\documentstyle[preprint,aps]{revtex}
\draft
\begin{document}
\title{Excitation functions in central Au+Au collisions\\
from SIS/GSI to AGS/Brookhaven}
\bigskip
\author{Bao-An Li and C.M. Ko}
\address{Cyclotron Institute and Physics Department,\\ 
Texas A\&M University, College Station, TX 77843}
\maketitle
 
\begin{abstract}
Using the relativistic transport 
model (ART), we predict the energy dependence of the 
stopping power, maximum baryon and energy densities,
the population of resonance matter as well as the strength of the 
transverse and radial flow for central Au+Au reactions at beam 
momentum from 2 to 12 GeV/c available at Brookhaven's AGS. 
The maximum baryon and energy densities are further compared to the 
predictions of relativistic hydrodynamics assuming the formation of 
shock waves. We also discuss the Fermi-Landau 
scaling of the pion multiplicity in these reactions. 
\\
{\bf PACS number(s): 25.75.+r} 
\end{abstract}
%\narrowtext
\newpage
\section{Motivation}
Experiments at Brookhaven's AGS have already revealed much interesting 
new physics during the last few years\cite{mit,qmatter94,qmatter95}. 
In the next few years several important experiments will be carried out 
at AGS to further study the properties of hot and dense hadronic matter, 
and to search for the signals for chiral symmetry restoration and/or 
Quark-Gluon-Plasma (QGP) formation\cite{guide}. Among the proposed experiments
we are particularly interested in those by the E895 and E866 collaboration
which will carry out extensive systematic studies at several beam momenta 
between 2 and 12 GeV/c using a Au beam\cite{eosn,ogilvie}. These 
experiments are interesting and important as it has been predicted 
based on hydrodynamical models that the QGP phase transition may 
occur in heavy-ion collisions at energies between $E_{lab}/A$=
2-10 GeV\cite{rischke,glen}. 
Also, theoretical studies 
(e.\ g.\ \cite{brown1,brown2,kaplan,hatsuda,asakawa}) 
have shown that the properties of hadrons, such as their masses and 
decay widths, may be modified in hot dense hadronic matter
as a result of the partial restoration of chiral symmetry.
Since the maximum baryon density achievable 
in heavy ion collisions at AGS energies varies from about 
$2\rho_0$ to $8\rho_0$ these experiments provide also a good opportunity 
to study the change of medium effects with density.
Comparisons between experimental and theoretical studies in this energy range 
are therefore very useful in searching for the signals for 
chiral symmetry restoration and/or baryon-rich QGP formation as well as 
testing various other predictions\cite{gyu}. In addition, 
how the reaction dynamics and the mechanism for particle production 
change in this energy range are also interesting by themselves.
  
In this paper, we present and discuss our predictions
on the excitation functions for a number of important quantities 
about the reaction dynamics and experimental observables 
in central Au+Au reactions using a relativistic transport model (ART 1.0). 
In particular, we examine excitation functions of the stopping power, 
maximum baryon and energy densities achievable, the population of 
resonance matter as well as the strength of the transverse and radial 
flow. The calculated maximum baryan and energy densities are further
compared with predictions of the relativistic hydrodynamics assuming
the formation of shock waves. We also discuss the 
excitation function of the pion multiplicity in terms of the
Fermi-Landau scaling variable.

This work is a continuation of our recent efforts in studying 
systematically the reaction dynamics, particle production mechanisms, the
properties of superdense hadronic matter\cite{art1,art2,art3}, and the
signatures of chiral symmetry restoration and QGP formation at AGS 
energies using the relativistic transport model (ART). 
Details of the transport model has been given in 
ref.\cite{art1}, here we will only give a brief summary of the 
model in section II. In section III we present and discuss our results.
A summary will be given at the end.    

\section{A brief summary about the model ART 1.0}
The model ART 1.0 is based on the very successful BUU transport 
model (e.g. \cite{bertsch,li91a,li91b}) for intermediate energy heavy-ion 
collisions by including more baryon and meson resonances as well 
as the interactions among them. More specifically, we have included in 
ART 1.0 the following baryons: $N,~\Delta(1232),~N^{*}(1440),~N^{*}(1535),
~\Lambda,~\Sigma$, and mesons: $\pi,~\rho,~\omega,~\eta,~K$, 
as well as their explicit isospin degrees of freedom. Both elastic and 
inelastic collisions among most of these
particles are included as best as we can by using as inputs 
the experimental data from hadron-hadron collisions.
However, more than 200 reaction channels 
are listed in the CERN data book\cite{data1} for the nucleon-nucleon and 
pion-nucleon collisions. We certainly have not fully incorporated all 
these channels. Instead, most of inelastic hadron-hadron collisions 
are modeled through the formation of baryon and meson resonances.
Although we only included explicitly the three baryon resonances 
$(\Delta(1232), N^*(1440)$ and $ N^*(1535)$), effects of heavier
baryon resonances are partially included by using 
meson+baryon cross sections calculated from the implicit formation 
of these heavier resonances with masses up to 2 GeV. 
We have also included in the model an optional, self-consistent 
mean field for baryons. 
 
Using this model we have recently studied in refs.\cite{art1,art2,art3} 
heavy-ion reactions at beam energies of 
about 12 GeV/nucleon. In particular, we have explored
effects of nuclear mean field on the collision dynamics, the 
volume and lifetime of the superdense hadronic matter, single particle 
observables as well as the radial and transverse collective 
flow of nucleons and pions. It has been found that the mean field significantly 
reduces the maximum energy and baryon densities of the superdense 
hadronic matter formed in the collision. On the other hand, inclusive, 
single particle observables are found to depend on the equation of state 
only at the level of 20\%. The mean field also affects significantly 
the transverse collective flow of baryons.
In particular, a soft nuclear equation of state increases the flow 
parameter by a factor of 2.5 compared to that from pure cascade calculations.

It is important to stress that ART is a purely hadronic model
and its validity naturally breaks down when the QGP 
is formed in the collision. However, results from hadronic transport 
models are useful as they provide the necessary background 
against which one can identify new physics phenomena due to, for example, 
chiral symmetry restoration and/or the formation of QGP. 
It is in this sense we present and discuss in the following
our model predictions. 

\section{Results and discussions}
We are particularly interested in the beam energy dependence
of several important quantities which may be related to the QGP phase 
transition and/or the chiral symmetry restoration in central Au+Au reactions. 
For some quantities we also compare 
predictions using the cascade mode of ART 1.0 and a soft nuclear equation of
state corresponding to a compressibility of 200 MeV. 
We restrict our discussions to the beam momentum range of 2 to 12 GeV/c
as systematic studies of heavy-ion collisions at 
lower energies have previously been studied extensively during the last decade 
by many groups using various transport models 
(for a review see, e.g., \cite{bauer}). In particular, 
one can find in ref.\ \cite{li94} a study on the beam energy 
dependence of Au+Au collisions using an early version of the present model.

\subsection{Beam energy dependence of compression and stopping power}
In this section, we shall find the maximum baryon and energy densities
reached in heavy-ion collisions in the energy range available at AGS
and compare them with the transition densities expected for the QGP.
We also would like to find the lowest beam 
energy at which the QGP transition may occur.

We first recall the notations used for specifying the compression. 
We use the standard test particle method\cite{wong82}
to calculate the global baryon density matrix $\rho_{g}^b(I_x,I_y,I_z,t)$ 
and energy density matrix $e_{g}(I_x,I_y,I_z,t)$ in the 
nucleus-nucleus c.m. frame on a lattice of size $40-40-48$ 
with a volume of 1 ${\rm fm}^3$ for each cell. The compression is then given
by the local baryon and energy density matrixes
obtained via $\rho_{l}^b(I_x,I_y,I_z,t)=\rho_{g}^b(I_x,I_y,I_z,t)/\gamma$, 
and $e_{l}(I_x,I_y,I_z,t)=e_{g}(I_z,I_y,I_z,t)/\gamma^2$,
where $\gamma$ is the Lorentz factor for each cell, i.e.

\begin{equation}
\gamma(I_x,I_y,I_z,t)=\sum_iE_i/\sqrt{(\sum_{i}E_i)^2-(\sum_i\vec{p_i})^2},
\end{equation}
with $i$ running over all particles in the cell.

To find the beam energy at which the maximum compression is reached, 
we examine the time dependence of the central baryon and energy densities
$(\rho/\rho_0)_{cent} \equiv \rho_{l}^b(0,0,0,t)$ and $(e_{l})_{cent} 
\equiv e_{l}(0,0,0,t)$ for Au+Au collisions at beam momenta from 
2 to 12 GeV/c and an impact parameter of 2 fm corresponding to typical 
central collisions. The results are shown in Fig.\ 1 and Fig.\ 2 for 
the case of pure cascade and the case of using the soft equation of state,
respectively. We notice that the maximum compression is reached at
the highest beam energy available at the AGS. Although the 
increase of the maximum density becomes slower starting 
at a beam momentum of about 8 GeV/c, it obviously does not 
saturate even up to 12 GeV/c. We thus expect that even higher densities 
can be reached in heavy-ion collisions at energies above the AGS energies.
This observation is useful for considerations of future experimental 
programs, such as those at the GSI\cite{metag}. 
Since the QGP phase transition energy density has been estimated to be about 
2 ${\rm GeV/fm^3}$\cite{wong95,kapusta1,kapusta2}, our calculations from
both the cascade and the transport model with the soft equation of state 
indicate that it can already occur at beam momenta 
above about 8 GeV/c. 

Relativistic hydrodynamics has been quite successful in studying many
aspects of relativistic heavy-ion collisions (for a recent review, see, 
e.g. \cite{dan1,dan2}). One aspect which is particularly interesting to us
in the present study is the possibility of forming shock waves in heavy-ion 
collisions (e.g., \cite{amsden}). Within the framework of relativistic 
hydrodynamics it was predicted that the shock waves are nearly 
completely developed in heavy-ion collisions at beam energies higher 
than about 400 MeV/nucleon\cite{dani84}. Furthermore, one can 
estimate the maximum compression behind the shock fronts
by studying the propogation dynamics of the shock waves.
It is therefore interesting to compare the predictions on the maximum 
compression from the transport model and the relativistic 
hydrodynamics model. Here we perform such a comparison for head-on 
collisions of Au+Au in the beam momentum range of 2 to 12 GeV/c. 

The maximum compression attainable in a heavy-ion collision 
can be obtained by either solving the Rankine-Hugoniot (RH)
equation\cite{taub} (e.g., in refs.\cite{nix,dan1,glen89,dani95}) 
or directly seeking for approximate solutions of the hydrodynamical equations 
(e.g., in refs.\cite{gale,ornik}). Here we solve numerically the RH 
equation using the approach detailed in refs.\ \cite{nix,dan1}.
The maximum baryon density $\rho$ satisfies the equation
\begin{equation}\label{shock}
f(\rho)(n-\gamma)-e_0n(B\gamma^2+\gamma^2-1-B\gamma n)=0,
\end{equation}    
where $n \equiv \rho/\rho_0$ and $e_0$ is the energy density of nuclear
matter ahead of the shock front. $B$ is the ratio of the thermal pressure
and the energy density. For a detailed discussion about effects of the
parameter $B$ on the maximum compression we refer the reader to section 2.4
of ref.\ \cite{dan1}. We found that a value of $B=2/3$ corresponding to a
non-relativistic gas can best describe our results from the transport model.
In the above, $f(\rho)$ is defined as
\begin{equation}
f(\rho)\equiv \rho^2d(\frac{e_0(\rho)}{\rho})/d\rho-Be_0(\rho),
\end{equation}
where $e_0(\rho)$ is the energy density at zero temperature. 
For a soft equation of state as used in the transport model, we have
\begin{equation}
e_0(\rho)/\rho=m_N+\frac{a}{2}\frac{\rho}{\rho_0}+\frac{b}{1+\sigma}(\frac{\rho}
{\rho_0})^{\sigma}+\frac{3}{5}E_f(\frac{\rho}{\rho_0})^{2/3}
\end{equation}
and 
\begin{equation}
f(\rho)=\rho_0\left(-\frac{2}{3}m_Nn+\frac{a}{6}n^2+\frac{b(\sigma-2/3)}
{1+\sigma}n^{\sigma+1}\right),
\end{equation}
where $a=-358.1$ MeV, $b=304.8$ MeV, $\sigma=7/6$,
$E_f$ is the Fermi energy in the ground state of nuclear matter 
and $m_N$ is the nucleon mass. 

To compare the prediction of relativistic hydrodynamics 
with that from our cascade calculations, we set $a=b=0$ and obtain the
analytical solution\cite{nix}
\begin{equation}
n=\frac{5\gamma^2-2\gamma-3}{2(\gamma-1)}
=\frac{1}{2}(5\gamma+3).
\end{equation}
The maximum baryon density and the maximum energy density obtained via 
$e_{max}=e_0\gamma\rho$ using the above formula are shown by the dotted lines 
in Fig.\ 3. The solid lines are obtained by solving numerically 
Eq. (\ref{shock}) using an iterative method for the soft equation of state. 
Results from the cascade and the 
transport model with the soft equation of state are shown by the open and
filled circles. We are amased to see that 
the predictions from the two calculations agree so well, especially 
at beam momenta below 9 GeV/c/nucleon. For head-on collisions of Au+Au 
at beam energies below 2 GeV/nucleon, Danielewicz\cite{dani95} has also 
compared predictions on the maximum compression using both the hadrodynamics 
model and the BUU model. He has found that the two results agree very well 
at beam energies above about 1.0 GeV/nucleon while at lower energies the 
hydrodynamics model predicts a higher compression. Our results are in 
agreement with his, and, furthermore, our study has extended the 
comparison to higher energies. We notice that there are some small 
discrepancies between predictions of our transport model and the hydrodynamics 
model at beam momenta higher than about 10 GeV/c.
These discrepancies are, however, expected. First, the system under 
study deviates from a non-relativistic gas at increasingly higher energies, 
and consequently the parameter $B$ will become 
smaller than 2/3. In fact, the approximate expressions for 
the maximum baryon density and the maximum energy density derived for an 
ultra-relativistic ideal gas in ref.\cite{ornik} using the relativistic 
hydrodynamics roughly reproduce our transport model calculations at 
high energies but underpredict significantly those at low energies. 
Second, conditions for applying the hydrodynamics, such as the local 
thermal equilibrium, may not be completely satisfied in transport 
model calculations at high energies. 
Another factor may also contribute to the difference is the finite size of 
lattice cells used to calculate the baryon and energy densities in the 
transport model.

The stopping power can be measured by the rapidity distribution
of nucleons. Previous experiments at AGS for central collisions of Au+Au 
at 12 GeV/c have shown nearly complete stopping\cite{e866n} and can be well
reproduced by our model calculations as shown in ref.\ \cite{art1}.
In Fig.\ 4, we show the rapidity distribution of nucleons at 6 different 
beam momenta between 2 and 12 GeV/c for the reaction of Au+Au 
at an impact parameter of 2 fm. The results are obtained with 
the soft equation of state. To gain information 
about the stopping dynamics, we further show the time evolution of 
the rapidity distribution from t=2 fm/c to 20 fm/c at a time 
interval of 2 fm/c. It is seen that a high degree of stopping is 
reached at all beam energies. The time to reach the stable rapidity 
distribution decreases rapidly as the beam energy increases. This 
can be understood in terms of the reduced reaction time, which scales 
approximately with the inverse of the Lorentz $\gamma$ factor and 
increased baryon-baryon collision rates with increasing beam momentum 
as shown in Fig.\ 5

\subsection{Beam energy dependence of the transverse and radial flow}
The transverse and radial flow of various hadrons in heavy-ion collisions 
at both low (NSCL/MSU, SIS/GSI and Bevelac/Berkeley) and 
high (AGS and CERN) energies have been studied extensively during the
last few years. One of the motivations of the 
E895 and E866 experiments is to study the strength of collective 
flows as the beam energy increases from the highest energy available 
at SIS/GSI and/or Bevelac/Berkeley to that at AGS/Brookhaven. 
The experimental results will be useful for testing various 
theoretical predictions. For example, recent quark-gluon-string model 
calculations have shown that the formation of a quark-gluon-plasma 
drastically reduces the transverse collective flow\cite{qgsm}. Also, 
it has been shown in a relativistic hydrodynamical model that the 
strength of collective flow should show a minimum at about 6 GeV/nucleon when 
the QGP transition occurs\cite{rischke95}. 

To study the transverse flow we use the standard 
method of Danielewicz and Odyniec\cite{dani}, i.e., 
analyzing the average transverse momentum in the reaction 
plane as a function of rapidity. We then extract the flow parameter
$F\equiv (dp_x/dy)_{y=0}$ and study its dependence on the beam energy.
Results of this analysis from both the cascade and the transport
model with the soft equation of state are shown in Fig.\ 6 for the 
reaction of Au+Au at an impact parameter of 2 fm. 
Two interesting features are worth of mentioning. First, in both cases 
the transverse flow parameter $F$ decreases quickly as the beam momentum 
increases from 2 to about 7 GeV/c, and it then more or less saturates. 
Second, the predicted flow parameter from the cascade is about a factor 
of two smaller than that using the soft equation of state in the whole 
energy range. It is well-known from experiments at Bevalac and SIS/GSI that the
strength of transverse flow increases with the beam energy up to about
1 GeV/nucleon in the reaction of Au+Au\cite{partlan}. There are, however,
still some discrepancies among data from different experiments as 
to whether the flow saturates or decreases above about 1 GeV/nucleon 
for the reaction of Au+Au. To understand the decrease in the strength 
of the transverse flow as the beam momentum increases above about 2 GeV/c,
we examine schematically the maximum flow parameter $F_{max}$ as a 
function of the beam momentum. For the purpose of qualitative discussions 
we assume that contributions to the flow parameter from nucleon-nucleon 
collisions and the mean field are additive,
i.e. 
\begin{equation}\label{flow}
F_{max}\approx (\langle\Delta p_x\rangle_{c})
/y_{cm}+(\int_{0}^{t_r}F_x dt)/y_{cm}.
\end{equation}
in the above $\langle \Delta p_x\rangle_c$ is the 
net transverse momentum generated by
nucleon-nucleon collisions and $F_x$ is the force acting on nucleons along 
the $+x$ direction in the reaction plane; $t_r\propto 1/\gamma$ is the 
reaction time. From the pure cascade calculations shown in Fig.\ 6 it is seen 
that the contribution from collisions decreases 
with increasing beam energy. This is mainly due to the rapidly decreasing 
reaction time $t_r$ during which the thermal pressure creates a sideward 
deflection in the reaction plane. The second term depends on both 
the reaction time and the density gradient which is expected to increase with
increasing beam energy. From the difference of the two calculations 
shown in Fig.\ 6 it is seen that the second term of Eq.\ (\ref{flow}) 
is almost a constant. Since both the reaction time $t_r$ 
and the center of mass rapidity $y_{cm}$ reduce the second term 
as the beam momentum increases, the above observation indicates that the 
force $F_x$ due to the mean field increases as a result of the larger 
density gradients achieved at higher energies.
The decrease of the flow parameter $F$ with increasing 
beam energy is thus a result of the reaction dynamics. 
Furthermore, one can conclude that 
the competing effects from the mean field and the reaction time result
in a peak in the strength of the transverse flow at beam momentum 
around 2 GeV/c. The experiments by the EOS/E895 collaboration 
in measuring the flow parameter from Bevalac energies up to the 
AGS energies are therefore very useful in understanding the effect 
of collision dynamics on collective flows.
  
To further understand how the transverse flow is generated during the 
reaction, we compare in Fig.\ 7 the time evolution of the flow
parameter $F$ at beam momenta of 2 and 12 GeV/c.
It is seen that $F$ saturates after about 17 and 10 fm/c, 
respectively, for the reaction at 2 and 12 GeV/c. It is interesting to
notice that at the beginning of the reaction at 2 GeV/c  
there is a weak antiflow due to the attractive interaction
between particles on the two surfaces. However, it eventually changes 
to a large positive value in agreement with our discussions above.

The strength of the radial flow is characterized by the average radial 
flow velocity $\beta_r$ defined as 
\begin{equation}
\beta_r\equiv \frac{1}{N}\sum_i^{N}\frac{\vec{p_i}}
{E_i}\cdot \frac{\vec{r_i}}{r_i},
\end{equation} 
where the summation is over all test particles in the system.
The value of $\beta_r$ at t=20 fm/c is shown in Fig.\ 8 for both the
cascade and the transport model with the soft equation of state. We 
see that the radial flow velocity increases quickly and saturates
at a value of about 0.78 c. The similar results from the two cases 
further indicate that the radial flow is mainly determined by the 
thermal pressure rather than the potential pressure. Since the thermal 
pressure increases with increasing beam energy as discussed above,
the radial flow increases thus with the beam energy.  

The larger average radial flow velocity obtained from the transport model
than those extracted experimentally from heavy-ion collisions 
at both Bevalac and AGS energies deserves some discussions.
Experimentally the average radial flow velocity is usually 
determined by fitting the particle spectra using a thermal source of 
temperature $T$ boosted by an azimuthally symmetric flow 
velocity $\beta$ which is taken to be proporntial to certain power of the 
radial coordinate $\beta(r)\propto r^{\alpha}$.
However, it was pointed out recently
that it was not possible to accurately extract the true temperature and 
the correct average radial flow velocity using this procedure\cite{kon}. 
Since the probability for a particle to receive a 
radial flow velocity $\beta$ is 
\begin{equation}
dP/d\beta\propto \rho(r)(\frac{d\beta}{dr})^{-1},
\end{equation} 
one needs to know not only the flow profile $\beta(r)$ but also 
the density distribution $\rho(r)$ which is certainly 
not uniform as usually assumed.
In fact, the extracted temperature assuming a uniform density 
is found to be much higher (e.g., a factor of about 2 for the
reaction of Au+Au at $E_{beam}/A$=150 MeV) than the chemical temperature 
extracted by using microscopic models\cite{kon}. The average 
radial flow velocity extracted experimentally is therefore 
correspondingly smaller.   
 
\subsection{Beam energy dependence of the abundance of resonance matter}
The possible formation of a resonance matter 
at SIS/GSI and/or AGS energies has been an interesting 
topic for many discussions (e.g.\, \cite{metag93,li94,hofmann}).
Here we examine the energy dependence of the creation and decay of 
baryon resonances in central Au+Au collisions at $p_{beam}/A$ from
2 to 12 GeV/c. In Fig.\ 9, we show the relative abundance of 
baryon resonances $R_1\equiv (N_{\Delta}+N_{N^*})/(A_t+A_p)$ (upper window) and 
$R_2\equiv (N_{\Delta}+N_{N^*})/(N_{\Delta}+N_{N^*}+N_{\pi})$ (lower window)
as a function of time and beam momentum. First, it is observed that the maximum
resonance/baryon ratio increases from about 20\% to 45\% as the beam
momentum increases from 2 to 12 GeV/c. This corresponds to an increase of
the resonance density from about $0.6\rho_0$ to $3\rho_0$. Of course, 
in a small volume around the center of the participant region 
the ratio is even higher. From the lower window, we see that the ratio $R_2$
in the expansion stage decreases as the beam energy increases. 
This is understandable from the pion production dynamics. 
At lower energies available at SIS/GSI, pions are 
almost completely produced through the resonances $\Delta(1232)$ and 
$N^{*}(1440)$. Moreover, on the average the resonances 
created at SIS/GSI energies have relatively lower masses 
and therefore longer lifetimes. As the beam energy increases not only 
the production of direct pions and meson resonances become increasingly 
important but also the created baryon resonances have heavier masses and 
correspondingly shorter lifetimes. Consequently, $R_2$ decreases
with increasing beam momentum.
 
It is important to point out here that one should not compare directly 
the calculated resonance/baryon ratio at any particular time with 
the ratio extracted experimentally by studying $\pi+N$ correlations, 
such as done by the E814 collaboration\cite{e814}, 
as calculations from both our model and the
recent quark-gluon string model indicate that hadrons are continuously 
emitted during the whole reaction process\cite{csernai}. Since pions 
and nucleons originating from the decay of baryon resonances near the 
surface of the reaction system can escape without any further interaction,
the number of resonances extracted from the experiments
therefore reflects the time-integrated emission of resonances in the reaction.

\subsection{Fermi-Landau scaling of the pion multiplicity}
Pions are the most copiously produced particles in relativistic heavy-ion
collisions. It was predicted that the existence of a mixed phase of quarks
and hadronic matter might result in a plateau-like behaviour of the excitation 
function of the pion multiplicity in central heavy-ion 
collisions\cite{gro1,gro2}. Further, the creation of a large entropy due to 
the formation of the QGP was expected to increase the average 
pion multiplicity\cite{gaz} as the beam energy increases across 
the transition energy, leading to a bump or shoulder in the pion 
multiplicity distribution\cite{kapusta2}.
We study in the following the average pion multiplicity 
as a function of the beam momentum.
In both Fermi's statistical model for particle production at high 
energies\cite{fermi} and Landau's relativistic adiabatic hydrodynamics
model\cite{landau}, the average pion multiplicity per participant nucleon
has been found to scale with the function $F_{FL}$ 
\begin{equation}
\frac{\langle \pi\rangle}{\langle N_p\rangle }\propto F_{FL}\equiv 
\frac{(\sqrt{s}_{NN}-2m_N)^{3/4}}{(\sqrt{s}_{NN})^{1/4}},
\end{equation} 
where $\sqrt{s}_{NN}$ is the center-of-mass energy of a 
nucleon-nucleon collision. Large deviations from this scaling 
might thus indicate the occurrence of exotic processes\cite{marek}. 
To take into account the threshold behaviour, Gazedzicki and R\"ohrich 
have recently modified the scaling function to the following form
\begin{equation}
F_{NN}\equiv 
\frac{(\sqrt{s}_{NN}-2m_N-m_{\pi})^{3/4}}{(\sqrt{s}_{NN})^{1/4}},
\end{equation} 
and investigated the scaling behaviour in both nucleon-nucleon (NN) and 
central nucleus-nucleus (AA) collisions from Bevalac energies up to CERN
energies\cite{gaz}. Their extracted $\langle \pi\rangle 
/\langle N_p\rangle$ data for NN and AA collisions 
are shown in Fig.\ 10 by open squares and filled circles, respectively.
The first three points for AA collisions are from the Bevalac data, 
the two points at $F_{NN}\approx 0.9$ are from the Dubna data and the
highest two points are extracted from the AGS data. Although 
several extrapolations and assumptions about the centrality
have been used in extracting the experimental information,
it is interesting to see that the pion multiplicity in both NN and AA
collisions scales with the Fermi-Landau variable $F_{NN}$.
The large difference between the two cases is probably due to 
the large pion reabsorption in AA collisions.
Our predictions from the transport model are shown by the open circles. 
They seem to scale with $F_{NN}$ and are actually very close to the extracted 
data for AA. Since it is still not easy to pin down the origin of the 
small difference between the calculated results and the extracted data 
for AA, more data between $F_{NN}$=0.4 and 1.7 to be available
from the E895 and E866 collaboration will certainly enable us to 
make a more detailed study on the scaling behaviour of the pion multiplicity.
     
\section{Summary}
Based on the relativistic transport model (ART 1.0) 
we have studied several excitation functions in central Au+Au reactions
at beam momenta from 2 to 12 GeV/c. We have found that the maximum
energy density increases with the beam energy and becomes higher than
the estimated QGP transition density at about 8 GeV/c. 
The calculated maximum baryon and energy densities are in good agreement
with predictions of the relativistic hydrodynamics assuming the formation
of shock waves in heavy-ion collisions. A high degree of stopping can be 
achieved in the whole beam momentum range. 
We have predicted that starting at beam momentum of about 
2 GeV/c the strength of the nucleon transverse flow 
decreases but then saturates when the beam momentum is higher than 
about 7 Gev/c. Also the strength of the radial flow is found to first 
increase as a function of beam momentum and then saturates. Furthermore, 
we have discussed the excitation function for the relative abundance of the 
resonance matter. The excitation function for the
pion multiplicity has also been discussed in terms of the Fermi-Landau 
scaling variable. We have found that the pion multiplicity indeed 
scales with the Fermi-Landau scaling variable in the whole energy 
range studied here. The calculated pion multiplicities are very close 
to the existing data. These results can be used as a baseline in 
searching for new physics phenomena that may appear due to  
chiral symmetry restoration and/or quark-gluon-plasma formation
in heavy-ion collisions at Brookhaven's AGS.

\section{Acknowledgement}
We would like to thank V. Metag and H. Oeschler for their 
suggestion and encouragement of carrying out this study. 
Helpful discussions with W. Bauer, P. Danielewicz, G.Q. Li, S. Pratt 
and K.L. Wolf are gratefully appreciated. We would also like to thank 
C. Ogilvie, Fan Zhu and F. Videbaek for helpful comments on the model ART. 
This work was supported in part by NSF Grant No. PHY-9212209 and PHY-9509266.

\newpage 
\section*{Figure Captions}
\begin{description}
 
\item{\bf Fig.\ 1}\ \ \
		The central baryon (upper window) and energy (lower window)
		densities as functions of time and beam momentum for the
		reaction of Au+Au at an impact parameter of 2 fm using 
		the cascade mode of ART.
\item{\bf Fig.\ 2}\ \ \
		Same as Fig.\ 1 but using a soft equation of state with 
		a compressibility of 200 MeV.	
\item{\bf Fig.\ 3}\ \ \
		The beam momentum dependence of the maximum central 
		baryon (upper window) and energy (lower window)
		densities for the reaction of Au+Au at an impact parameter 
		of 2 fm. The dotted and solid lines are predictions 
		of the relativistic hydrodynamics.
\item{\bf Fig.\ 4}\ \ \
		The nucleon rapidity distribution as a function of 
		beam momentum and time for the collision of Au+Au at an impact 
		parameter of 2 fm.
\item{\bf Fig.\ 5}\ \ \
		The baryon-baryon collision rate as a function of time and beam 
		momentum for central collisions of Au+Au at an impact 
		parameter of 2 fm.
\item{\bf Fig.\ 6}\ \ \
		The transverse flow parameter as a function
		of beam momentum for the reaction of Au+Au at an impact
		parameter of 2 fm. 		
\item{\bf Fig.\ 7}\ \ \
		The transverse flow parameter of nucleons as a function
		of time for the reaction of Au+Au at an impact
		parameter of 2 fm and beam momentum of 2 and 12 GeV/c 
		respectively. 		
\item{\bf Fig.\ 8}\ \ \
		The average radial flow velocity as a function
		of beam momentum for the reaction of Au+Au at an impact
		parameter of 2 fm. 		
\item{\bf Fig.\ 9}\ \ \
		The relative abundance of baryon resonances
		with respect to the total number of baryons and pion-like
		particles as a function of time and beam momentum for 
		the reaction of Au+Au at an impact parameter of 2 fm. 		
\item{\bf Fig.\ 10}\ \ \
		The average pion multiplicity per participant nucleon 
		as a function of the Fermi-Landau scaling variable.
		The lines are drawn to guide the eye.
\end{description}
 
\end{document}